\documentclass[aps,pra,superscriptaddress,12pt,longbibliography]{revtex4-2}

\makeatletter
\renewcommand{\p@subsection}{}
\renewcommand{\p@subsubsection}{}
\makeatother

\usepackage{amsmath,amsfonts,amssymb,amsthm}
\usepackage{mathtools}
\usepackage{setspace}
\usepackage[normalem]{ulem} 
\usepackage{stmaryrd}       
\usepackage{dsfont}         
\usepackage{expdlist}
\usepackage{physics}
\usepackage{braket}
\usepackage{verbatim}
\usepackage{xspace}
\usepackage{pgf}
\usepackage{tikz}
\usetikzlibrary{arrows,automata}
\usetikzlibrary{shapes.geometric}
\usetikzlibrary{arrows.meta}
\usepackage[latin1]{inputenc}
\usepackage{verbatim}
\usepackage{textcomp} 

\usepackage{graphicx}
\usepackage{xcolor}
\usepackage{hyperref}
\hypersetup{pdfstartview=}
\usepackage{url}

\usepackage{alltt}
\usepackage{moreverb}

\usepackage{xr} 

\providecommand{\ignore}[1]{}

\newif\ifcmnt
\cmnttrue
\ifdefined\cmntsoff\cmntfalse\fi

\ifcmnt
    \providecommand{\aucmnt}[1]{#1}

\else
    \providecommand{\aucmnt}[1]{}

\fi

\newcommand{\Prob}{\mathbb{P}}

\newcommand{\cA}{\mathcal{A}}

\newcommand{\cL}{\mathcal{L}}

\newcommand{\cS}{\mathcal{S}}
\newcommand{\cT}{\mathcal{T}}

\newcommand{\cY}{\mathcal{Y}}

\numberwithin{equation}{section}

\newtheorem*{theorem*}{Theorem}

\newtheorem*{lemma*}{Lemma}

\newcommand{\rar}{\rightarrow}

\newcommand{\imp}{\implies}

\let\originalleft\left
\let\originalright\right
\renewcommand{\left}{\mathopen{}\mathclose\bgroup\originalleft}
\renewcommand{\right}{\aftergroup\egroup\originalright}

\DeclareMathOperator*{\argmax}{argmax}

\begin{document}

\title{Improving quantum state detection with adaptive sequential observations}

\author{Shawn Geller}
\affiliation{National Institute of Standards and Technology, Boulder, Colorado 80305, USA}
\affiliation{Department of Physics, University of Colorado, Boulder, Colorado 80309, USA}
\author{Daniel C. Cole}
\altaffiliation[Current address: ]{ColdQuanta, Inc., Boulder, Colorado 80301, USA}
\affiliation{National Institute of Standards and Technology, Boulder, Colorado 80305, USA}
\author{Scott Glancy}
\affiliation{National Institute of Standards and Technology, Boulder, Colorado 80305, USA}
\author{Emanuel Knill}
\affiliation{National Institute of Standards and Technology, Boulder, Colorado 80305, USA}
\affiliation{Center for Theory of Quantum Matter, University of Colorado, Boulder, Colorado 80309, USA}

\begin{abstract}
For many quantum systems intended for information processing, one
detects the logical state of a qubit by integrating a continuously
observed quantity over time. For example, ion and atom qubits are
typically measured by driving a cycling transition and counting the
number of photons observed from the resulting fluorescence. Instead of
recording only the total observed count in a fixed time interval, one can observe the photon
arrival times and get a state detection advantage by using the
temporal structure in a model such as a Hidden Markov Model.  We
study what further advantage may be achieved by
applying pulses to adaptively transform the state during the
observation.  We give a three-state example where adaptively chosen
transformations yield a clear advantage, and we compare performances
on an ion example, where we see improvements in some
regimes.  We provide a software package that can be used for
exploration of temporally resolved strategies with and without
adaptively chosen transformations. 
\end{abstract}

\maketitle

\section{Introduction}
Quantum information processing requires high fidelity single-shot
readout of states.  In a typical example, readout of the state of an ion or atom qubit is performed by observing the
fluorescence from driving a cycling transition. For a qubit whose states are superpositions of two atomic levels, a goal
is to distinguish between the two levels with the highest possible
fidelity.  A common way to define the fidelity is as the average
probability of correctly determining the level  when the levels are
prepared uniformly randomly. High fidelity readout is helpful in quantum
error-correction for syndrome measurements to minimize the
probability  of misidentifying errors. High fidelity readout can also significantly reduce the
number of measurements needed to characterize states or processes in
quantum tomography. The improvement can be substantial when the states
measured have high fidelity with respect to a target state
such as a Bell state~\cite{Tropp2010}. 

A standard readout method 
for `bright' and `dark' atomic levels (labeled $\ket{b}$ and $\ket{d}$
respectively)
is to drive a cycling
transition such that $\ket{b}$ fluoresces while 
$\ket{d}$ does not.  The emitted photons are detected with some
efficiency, and the output of the measurement is the number of photons
detected. The presence of background photons, unwanted transitions between states and the desire to have short observation times  prevent arbitrarily high fidelity
measurement. The atom is inferred to be in
the state $\ket{b}$  if the number of photons detected exceeds
a threshold, otherwise it is inferred to be
in the state $\ket{d}$. One strategy
to improve the readout fidelity is to use repetitive readout with
ancillary atoms as demonstrated in Ref.~\cite{Hume2007}.  A less
demanding strategy is to record and use the arrival time of photons as
 discussed in Ref.~\cite{Langer2006}.  
The arrival-time record can
  also be used with machine learning strategies to improve readout, as
  demonstrated in
  Ref.~\cite{Cappellaro2019,zihanguangcan2019,crain2019high,seif2018machine,magesan2015machine}. All these methods take
advantage of the fact that measurements are processes and can yield a
time-resolved record 
that contains
valuable
information about the initial logical state. 
While we focus on
atoms and fluorescence measurement, this situation is also common for other
systems being investigated for quantum information processing, such as
superconducting qubits~\cite{Gambetta2007}.

The measurement processes of interest are well described by hidden
Markov models (HMMs)\cite{Ephraim2002}. An HMM is a stochastic discrete-time process on
a finite number of ``hidden'' states with stochastic output. It is
characterized by a finite state space $\cal{S}$, an output space
$\cal{Y}$, a stochastic transition matrix $A$ that describes the
probabilities of transitioning from the current state to the next in a time
step, an output process matrix $B$ that determines the
probability of an output given the 
current
state, and an initial state distribution $\nu$ that determines the probability
of the HMM starting in a particular state. For applications to quantum measurement, the goal is to infer the
  starting state of the HMM from the sequence of outputs observed
  during the process.
  It is usually possible to infer the parameters of the HMM by fitting to experimental data using standard algorithms~\cite{Rabiner1989}, and one can assist such an algorithm by initializing it with parameters that are motivated by theory.
Once the parameters are known, statistical decision methods for HMMs can be applied to estimate the initial state, and this has
  been demonstrated in several recent works~\cite{schoelkopf2020,dubois2020,Hann2018,Curtis2020}. 

But it is possible to do even better  if
the systems of interest  permit sufficient control to
modify the steps during the measurement.  For atoms, the simplest such
modification is to apply pulses between
otherwise identical steps to permute the levels.  This can improve the measurement fidelity
or shorten average measurement times
by taking advantage of the fact that 
not all levels are equally distinguishable by their output fluorescence.
For example, two non-fluorescing levels are difficult to distinguish, but if
the output observed so far suggests that the atom was prepared in a non-fluorescing level, we can swap one of these
levels with a fluorescing one to learn which one was present initially,
provided the measurement process has not yet induced transitions between
levels.
  While in principle swaps can be applied in any system that permits such
  control, we expect them to be most useful for improving detection fidelity in systems where the time it takes
  to apply a swap is small compared to the lifetime of the system. 
  An example of such a system, which we discuss in this paper, is a $^9\text{Be}^+$ hyperfine qubit with microwave controls.

An improvement in measurement fidelity by applying 
a fixed  swap  was demonstrated
 in~\cite{Woelk2015,Hemmerling2012},
where
the authors use a single $\pi$ pulse to implement a level swap in the middle of the
readout cycle and postselect 
on the event
that the two halves of the readout cycle give different outputs. 
Strategies
that use postselection can provide conditional advantages in measurement
fidelity over those that do not, but we do not consider postselected measurement
fidelities here. In this work, rather than using only a fixed swap, we formalize and explore measurement processes with state permutations adaptively chosen during a single readout cycle based on the outputs observed so far in that cycle.

  To support  exploration of measurement processes with adaptively chosen state permutations, we developed a software
  package to simulate and optimize policies for choosing permutations \cite{github}.
 We consider two examples to illustrate and compare measurement strategies.  The first is a $^9\text{Be}^+$ ion, and the
  second is an idealized three-state example.  
    For these two examples, we computed the optimal policy for small numbers of
    steps, and  for the $^9\text{Be}^+$ example, we compared  the performance of a heuristic
    policy that is easier to compute than the optimal one. Improvements for the
  $^9\text{Be}^+$ example are moderate, but they
  illustrate what can be achieved with adaptive control policies in an
  experimentally relevant system.
    Improvements for the three-state example are more significant, showing that
    adaptive policies can be very useful in some cases.
  
  The heuristic policy is suboptimal for a range of parameters, demonstrating
  that  heuristic approaches can yield undesirable results.
  It is possible to formulate the problem in terms of partially observed Markov
  decision processes (POMDPs)~\cite{Astrom65}, which generalize HMMs by
  including output-dependent actions.
  In general, such problems are
  hard~\cite{papadimitriou1987complexity}, but approximate solutions
  exist~\cite{kurniawati2008sarsop}. Because the problems considered here are
  small, we did not need to apply approximate solutions, but applications of
  the proposed techniques to other systems with larger numbers of steps or
  larger output spaces will require approximation.

  In Sect.~\ref{sec:setup}, we describe the problem of initial state inference
  under repeated measurements, subject to adaptive control of the underlying
  states.
We discuss requirements for an experimental implementation of
the proposed protocol in Sect.~\ref{sec:expimp}.
  The two examples are introduced
  and analyzed in Sect.~\ref{sec:be9} and Sect.~\ref{sec:threestates}. We discuss the results and suggest extensions of this
  work in Sect.~\ref{sec:discussion}, and we offer concluding comments in
  Sect.~\ref{sec:conclusion}.  Mathematical formalisms, and technical descriptions
of policies and their computations are in the appendices.

    \section{Statement of problem}

\label{sec:setup}

Consider a generic atom being measured by observing fluorescence from
driving a cycling transition. We can divide the measurement into
equal-time intervals referred to as `steps'. For each step, we record
the number of photons detected. 
Many measurement configurations have the property that the system observed  can be treated as decohered in the measurement basis before each measurement step.
For atoms, the measurement basis is determined by
the atomic levels, and the effective decoherence is a consequence of
the dynamics of the levels and the nature of the driving light. Here we assume that the levels are non-degenerate, and each level can be treated as a single quantum state.

When decoherence between the levels does not arise naturally, in
many cases it is possible to enforce this decoherence and
eliminate memory of coherences between levels by active means such as
random phase changes or appropriately randomized pulses between steps.
Given the lack of phase memory, the action of
a step has a classical description. In particular, the initial quantum
state is a probability distribution over the levels $\mathcal{S}$ that constitute
the measurement basis elements. 
  The output and next level have a joint probability distribution that depends
  on the current level.

 Under these conditions the problem is reduced to a classical one, and we
describe the measurement dynamics with a Hidden Markov Model (HMM), which is a
discrete-time stochastic process.  From here on ``state'' refers to a classical
deterministic state, and we use probability distributions over states and
outputs to describe processes.  We describe the mathematics of HMMs in
App.~\ref{sec:repetitive}, and an exposition of how to describe measurement of a
$^9\text{Be}^+$ hyperfine qubit in terms of an HMM is given in
App.~\ref{sec:formalismexample}.

 The  description of a sequence of $n$
measurement steps requires the sequence of states
 $s^{n} =
(s_{1},\ldots, s_{n})$ and the sequence of outputs or observations
$y^{n} = (y_{1},\ldots, y_{n})$. The corresponding random variables
(RVs) are denoted by $S^{n}$ and $Y^{n}$. We use the usual convention
that RVs are denoted by capital letters and their values by the
corresponding lower-case letters
. For any sequence $x^{n}$, we use
$x^{j}$ to denote the initial subsequence that has $j$ elements, $x_1, \ldots,
x_j$.

After observing the outputs $y^n$, we make an inference of the 
  state occupied at the beginning of the first step, also called the initial state.
The random variable for the inferred initial state is denoted $\hat{S}_1$.
Using the HMM, we apply the maximum likelihood method  to infer
the initial state.
To characterize the performance of our inference method, we use the infidelity
$\Prob(S_1 \neq \hat{S}_1)$.
The symbol $\Prob$ denotes ``probability of'' and its argument is an event.
To compute the infidelity, we use the fact that the infidelity is the expectation value of the indicator function
of the event that the inference is incorrect, averaged over the prior $\Prob(s_1)$ and over the model
$\Prob(y^n|s_1)$ of the event that the inference is incorrect.  That is,
\begin{align}
  \Prob(S_1 \neq \hat{S}_1) &= \sum_{y^n, s_1}\mathbb{I}[s_1 \neq \hat{s}_1(y^n)]\Prob(y^n|s_1)
  \Prob(s_1),
  \label{eq:infidelityinterp}
\end{align}
where the indicator function $\mathbb{I}[E]\in\{0,1\}$
is $1$ if and only if $E$ is true. 
As discussed in App.~\ref{sec:initial},  maximum likelihood inference
minimizes the infidelity for the uniform prior.
Eq.~\ref{eq:infidelityinterp} can be computed exactly using the HMM if the number of steps is not too
large.

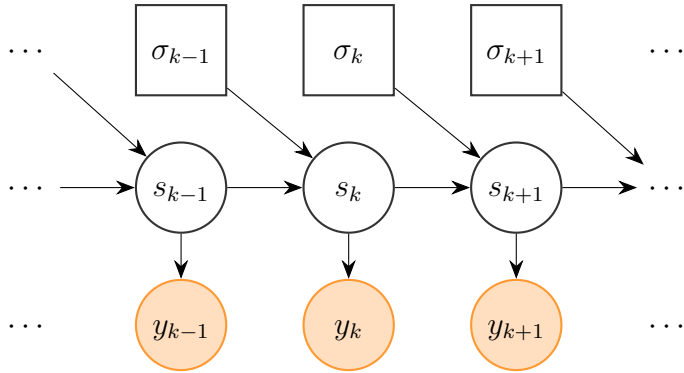
\begin{figure}[h]
\tikzstyle{action}=[rectangle, thick, minimum size=1.2cm, draw=black!80]
\tikzstyle{state}=[circle, thick, minimum size=1.2cm, draw=black!80]
\tikzstyle{measurement}=[circle, thick, minimum size=1.2cm, draw=orange!80, fill=orange!25]
\centering
\begin{tikzpicture}[text height=1.5ex,text depth=0.25ex]
\matrix[row sep=.6cm,column sep=1cm] {
      \node (sigma_k-2)         {$\cdots$};           &
      \node (sigma_k-1) [action] {$\sigma_{k-1}$}; &
      \node (sigma_k)   [action] {$\sigma_k$};     &
      \node (sigma_k+1) [action] {$\sigma_{k+1}$}; &
      \node (sigma_k+2)         {$\cdots$};           \\
      \node (s_k-2)         {$\cdots$};           &
      \node (s_k-1) [state] {$s_{k-1}$}; &
      \node (s_k)   [state] {$s_k$};     &
      \node (s_k+1) [state] {$s_{k+1}$}; &
      \node (s_k+2)         {$\cdots$};           \\
      \node (y_k-2)         {$\cdots$};           &
      \node (y_k-1) [measurement] {$y_{k-1}$}; &
      \node (y_k)   [measurement] {$y_k$};     &
      \node (y_k+1) [measurement] {$y_{k+1}$}; &
      \node (y_k+2)         {$\cdots$};           
      \\
  };
\path[->,>={Stealth[scale=1.5]}]
  (sigma_k-2) edge (s_k-1)
  (sigma_k-1) edge (s_k)
  (sigma_k) edge (s_k+1)
  (sigma_k+1) edge (s_k+2)

  (s_k-2) edge (s_k-1)	
  (s_k-1) edge node 
  {} (s_k)		
  (s_k)   edge node
  {} (s_k+1)	
  (s_k+1)   edge (s_k+2)	
 
  (s_k-1) edge[right] node {} (y_k-1)
  (s_k)   edge[right] node {} (y_k)
  (s_k+1) edge[right] node {} (y_k+1)

  ;
\end{tikzpicture}
\caption{A schematic diagram describing the problem considered
in this work. The arrows in the diagram indicate the
  dependencies between the different variables. The states $s_k$ are identified with the classical state of the experimental system
during  step $k$ of the measurement process,
 the outcomes $y_k$ are the
corresponding outcomes of each measurement step, and the permutations $\sigma_k$ of each state $s_k$ are chosen adaptively in each step.  The dependence of $\sigma_{k}$ on $y_{1},\ldots, y_{k-1}$ is not
depicted. This dependence is used in the experimenter's rule for choosing
permutations to apply. In this work, we study the problem of finding
$\sigma_{k}$'s that minimize the probability (Eq.~\ref{eq:infidelityinterp}) of incorrectly inferring the
initial state $s_1$.}
\label{fig:phmm}
\end{figure}

It is possible to modify the measurement sequence by using the data observed thus
far to act on the system using controls of the underlying states. 
We specialize to a set $\mathcal{A}$ of permutations of underlying states,
usually implemented with $\pi$-pulses between
the states.
As shown in Fig.~\ref{fig:phmm}, at each step, we  apply a permutation of the underlying states (this can be the identity permutation)
that depends on the data seen thus far.
A rule that determines which permutation to apply at each step is called a policy.
We would like to minimize the infidelity Eq.~\ref{eq:infidelityinterp} over the
space of all possible policies.
In general, computing the optimal policy requires an exponential time
calculation. This motivates the search for policies that are easier to
compute, but that are possibly suboptimal. One possible policy is to choose the permutation that maximizes the mutual information
between the initial state and the next $g$ outputs. We call this the minimum
posterior entropy heuristic. The mathematical descriptions of the HMM with
adaptive permutations, the optimal permutation policy, and the minimum
posterior entropy heuristic policy are given in
App.~\ref{sec:supadaptive}.

In Sect.~\ref{sec:be9} we compare the performance of the optimal policy 
and the minimum posterior entropy heuristic policy
for a certain set of permutations
to a policy that applies no
permutations in the context of a
$^9\text{Be}^+$ ion qubit under repeated fluorescence measurement. In
Sect.~\ref{sec:threestates}, we compare the performance of the optimal policy
to that of the trivial policy that does not apply permutations in an idealized
three state example.

\section{Prospects for experimental implementation}
\label{sec:expimp}

In this work, we focus on the $^9\text{Be}^+$ hyperfine qubit as
a motivating example. The fluorescence detection process in $^9\text{Be}^+$ can
be paused; by turning off the detection laser, the qubit retains its state
for a very long time. By turning off the laser, applying the
permutation pulse, then turning the laser back on, we can justify the assumption that
the permutations occur instantaneously. In other systems such as
superconducting qubits, the instantaneous action assumption does not hold. To
account for the finite duration of actions, one  can use the 
formalism of partially observable Markov decision
processes (POMDPs)~\cite{Astrom65} to account for the possibility that the
transition probabilities and the output probabilities directly depend on the
action.
The connection between the problem considered in this work with POMDPs is
discussed in App.~\ref{sec:supadaptive}.

The policies presented in this work are computed in advance. The advantage of
computing a policy in advance is that it can be turned into a look-up table, so that the
logic involved can  be implemented on a field-programmable gate array (FPGA). One downside is that if
the number of steps is too large, the look-up table can grow to the point that
it cannot be stored in memory. Another strategy is to compute a policy as the data comes in. Such strategies are not memory-limited, but have
the downside that typically floating-point calculations are involved, making
them difficult  to implement and run them in real time on an FPGA.

\section{Example: \texorpdfstring{$^9\text{Be}^+$}{9Be+} Hyperfine qubit}
\label{sec:be9}

\begin{figure}[h]
  \centering
    \def\svgwidth{.5\columnwidth}
    \scalebox{1}{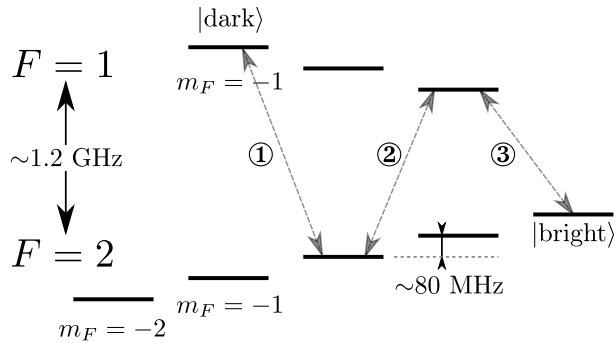}
    \caption{The ground state hyperfine manifold of the $^9\text{Be}^+$ ion in an applied magnetic field. The
dashed arrows show levels that differ by $\Delta F = \pm 1$, and $\Delta m_F = 0, \pm 1$. 
 Also indicated are the $\Ket{\text{dark}}$ and
$\Ket{\text{bright}}$ levels, which are the levels on which we assume a uniform
prior. The prior
assigns zero probability to all other levels. All unlabeled
levels fluoresce at the same low rate as the $\Ket{\text{dark}}$
level. Also indicated are the hyperfine splitting of approximately 1.2 GHz
between the $F=1$ and $F=2$ levels, and the Zeeman splitting of approximately
80 MHz at 0.0119 T. 
  By composing pairwise swaps along the arrows in the order shown, we can move
  the dark state to the bright state. This is the permutation $\tau$ considered
  in the main text. Reversing the order gives its inverse $\tau^{-1}$. When
  constructing policies, we allow the actions $\left\{ \tau, \tau^{-1},
\epsilon \right\}$ where $\epsilon$ is the identity permutation.
}
  \label{fig:be9levels}
\end{figure}

The $^9\text{Be}^+$ ground-state hyperfine qubit is measured by distinguishing
a level that fluoresces from ones that do not. The
level that
fluoresces does so by participating in a cycling transition, driven by
a detection laser. A cycling transition involves driving the internal
state to one that is outside the ground-state manifold, which in turn
immediately decays back to the state from which it came, emitting a
fluorescence photon in the process. Ideally, the other states remain
undisturbed in this process. In reality, imperfect polarization of the detection
laser can drive transitions from the bright state to dark states, while off-resonant driving leads
to transitions between other pairs of levels. 
Since we only consider the case of perfect polarization in this work,
if the ion ever enters the bright state, it remains there for the rest of the
measurement sequence.

 A level diagram of the $^9\text{Be}^+$ system is in Fig.~\ref{fig:be9levels}.  The fluorescing, `bright', level is $(F, m_F) = (2, 2)$,
and the level with the least probability of transitioning to $(2, 2)$ is
$(1, -1)$, so we use these two levels as possible initial states, with the uniform
prior $\Prob(s_1 = (2,2)) = \Prob(s_1 = (1, -1)) = 1/2$. 
The rate equations for this system are
derived in~\cite{paschke20179be}. 
The HMM parameters were obtained by
integrating the rate equations for a $^9\text{Be}^+$ qubit tuned to the
first-order magnetic-field-insensitive point for the $\Ket{F=1, m_F=1}
\leftrightarrow \Ket{F=2, m_F=0}$ transition~\cite{Langer2006,acton2005}.
The rate equations are parameterized 
by the background detected photon rate $\gamma_{\mathrm{bg}}$, the bright-state detected photon rate
$\gamma_{c}$,  the fractions of $\sigma^{-}$ and $\pi$ polarizations of the
detection laser, and the on-resonance saturation parameter for the cycling
transition $r$
that parameterizes the number of scattered photons per unit time.
To simplify the equations we assume that the polarization of the detection laser is perfectly $\sigma^+$.
We set the other parameters for the rate equation as follows:
$\gamma_c = $ 30 photons per 330 $\mu$s, $\gamma_{bg} = 0.06 \gamma_c$, and
$r=1/2$.

\begin{figure}[h]
  \centering
    \def\svgwidth{.85\columnwidth}
    \scalebox{1}{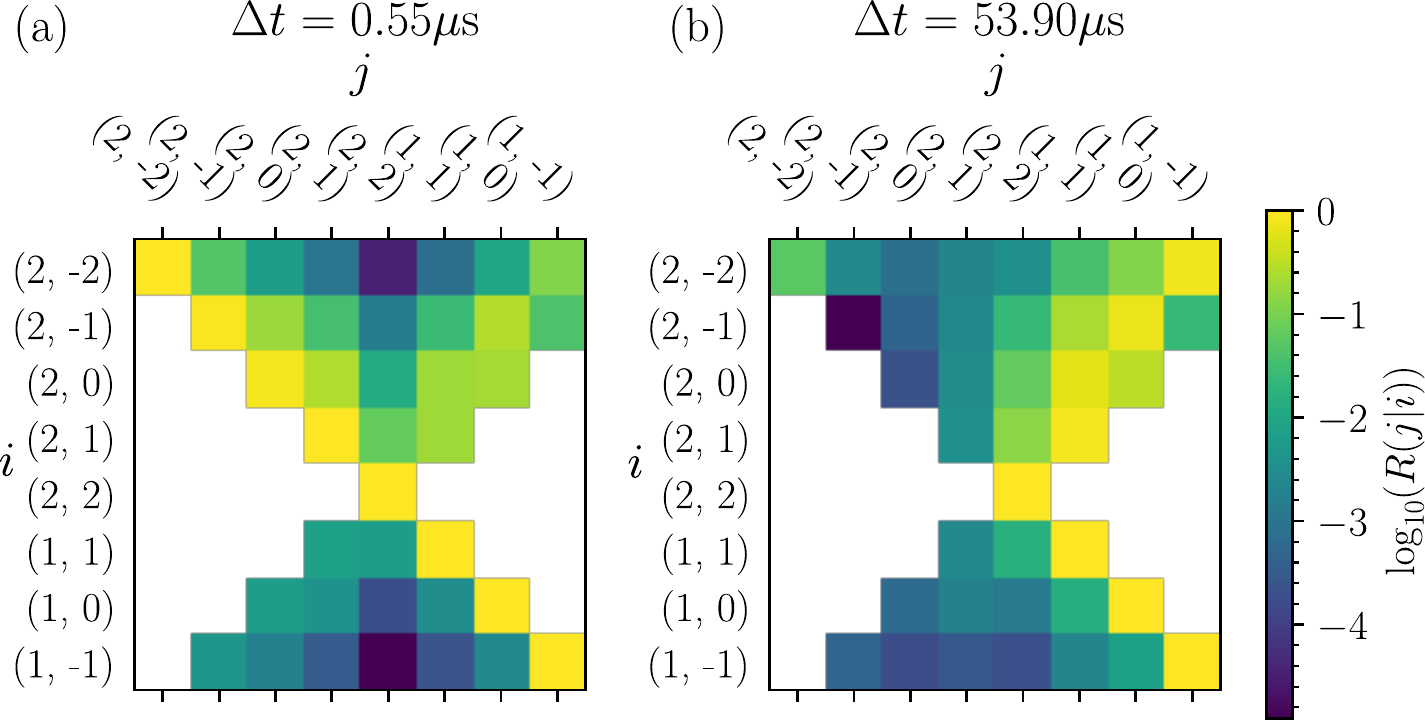}
  \caption{The entrywise $\log_{10}$ of the transition matrix between levels in the ground state hyperfine manifold of a $^9\text{Be}^+$ ion.
  If a transition is forbidden by the
assumption of perfect polarization, its entry $R(j|i)=0$ is shown in white.
  The rows
  correspond to the state before the transition, while the columns correspond
  to that after. The labels indexing the rows and
    columns of the matrix refer to $(F, m_F)$
  labels for the ion levels.  (a) shows the transition matrix at the smallest time per step
$\Delta t$ considered, while (b) shows that at the largest $\Delta t$
considered.}
  \label{fig:tmats}
\end{figure}

By integrating the rate equations for a time per step $\Delta t$, we derive
an HMM that describes the repeated measurement setup, as detailed in
App.~\ref{sec:formalismexample}.
We display the resulting transition matrices in
Fig.~\ref{fig:tmats}, for different values of  $\Delta t$ considered. 
We note that in Fig.~\ref{fig:tmats} (b), we observe large probabilities  of
transitioning between states. This is to be expected, as the detection
laser is assumed to have $\sigma^+$ polarization, which leads to
off-resonant driving of population to states with larger $m_F$.

\begin{figure}[h]
  \centering
    \def\svgwidth{.8\columnwidth}
    \scalebox{1}{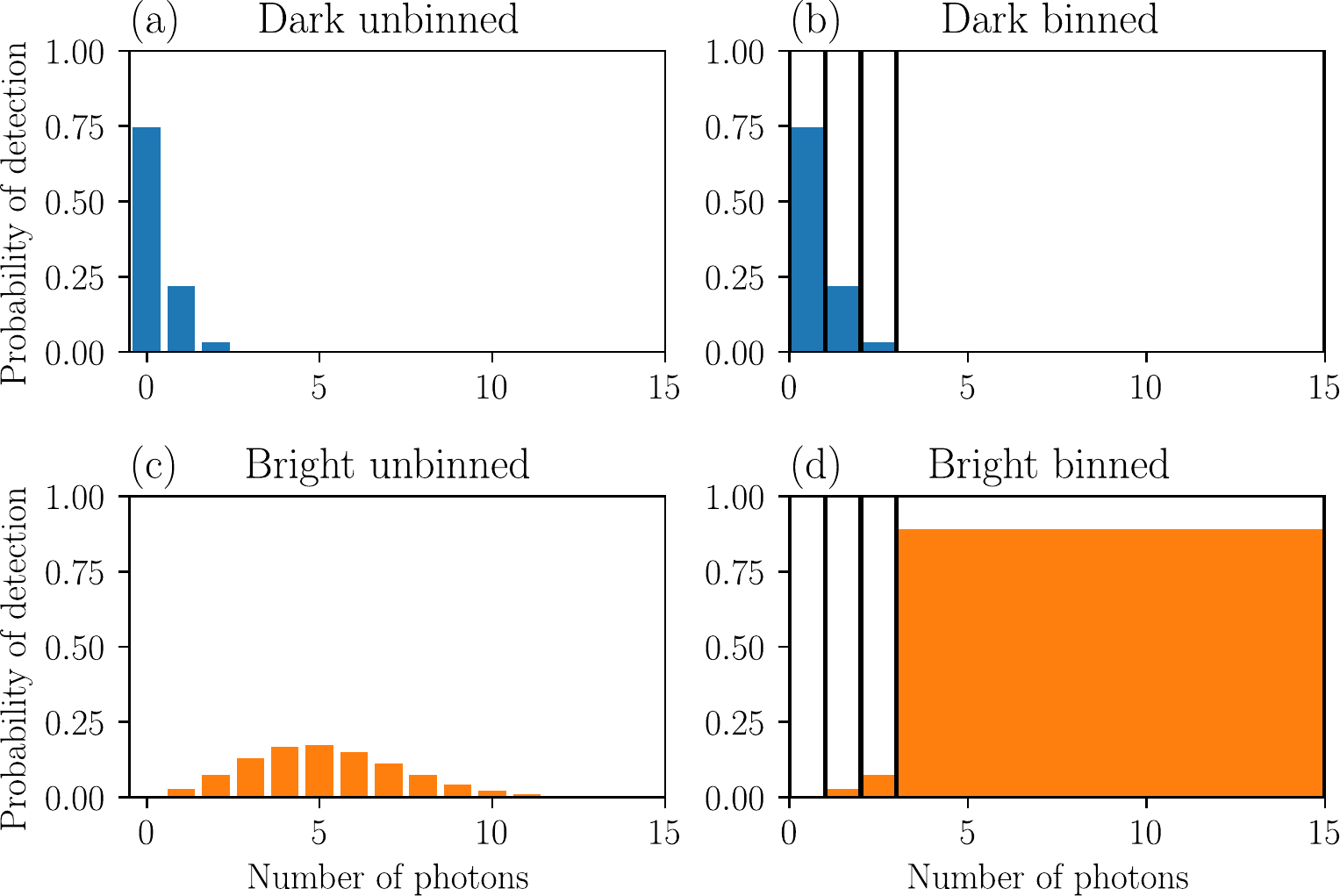}
  \caption{Bar charts of the probabilities of the various measurement outcomes,
    for a time per step $\Delta t$ of 53.9$\mu$s. This is the largest $\Delta t$
    that we consider in Fig.~\ref{fig:exact}.
  On the left in subfigures (a) and (c), we show the unbinned distributions for
collecting $n$ photons, conditioned on being in the state $(F, m_F) = (1, -1)$
and $(F, m_F) = (2, 2)$ respectively. These states are referred to as ``dark''
and ``bright'' in Fig.~\ref{fig:be9levels}, and the initial state distribution
is uniform over these two states. On the right in subfigures (b) and (d) are
the binned versions of the distributions shown in (a) and (c)
obtained by choosing non-uniform bins for the histogram
  counts of the number of photons observed.
The vertical black bars define the edges of the bins, and summing the
probabilities of outcomes that lie within a bin yields the probability of
observing that bin. 
}
  \label{fig:binned_histograms}
\end{figure}

Because the complexity of the calculations grows very rapidly with the number
of possible outcomes, it is expedient to 
  modify the output distributions to have fewer possible outputs. We reduce the
  possible outputs by identifying sets of numbers of photons collected in
  a step, and
  keeping only the information of which set the photon count lies in.
  A set of photon numbers is called a `bin'.
To select a set of bins, we minimized the infidelity over possible choices of bins, at $n = 6$ steps, when applying no
permutations, for $n_b = 4$ bins.  To reduce computation time, we restrict the search to bins
containing consecutive numbers of photons. 
  The optimization for selecting bins is presented formally in App.~\ref{sec:formalismexample}.
We show the resulting binned distributions in Fig.~\ref{fig:binned_histograms}
for the largest $\Delta t$ considered.
We performed this optimization separately for each $\Delta t$ we considered.  Once we find the
optimal binning,  we use
the same binning for all the different policies considered.

  The set of actions $\mathcal{A}$ used consists of the permutations  $\tau$, which sends the dark level
  to the bright level, its inverse $\tau^{-1}$, and $\epsilon$ the identity
  permutation. The permutation $\tau$ is a composition of three pairwise swaps
  as shown in Fig.~\ref{fig:be9levels}. The pairwise swaps that are used in
  creating $\tau$ are a subset of the transitions allowed by selection rules
  that are experimentally convenient to implement.
 While in principle we could add more actions to the set, we found  
that adding more actions does not appreciably improve measurement fidelity.
We assume that the error introduced by applying actions is negligible.
In the $^9\text{Be}^+$ system, the state of the system is very well preserved when the detection laser is turned off.
By assuming that the detection laser is turned off when the permutations are applied, we can treat the actions as effectively happening instantaneously.

We wish to analyze the effectiveness of using permutation actions in improving
measurement fidelity.  
  For $n=6$ steps, we are able to compute the optimal policy (see
  App.~\ref{sec:supadaptive}) using $n_b=4$ bins. However,
    the computational work required in computing the optimal policy grows as
    $O((|\mathcal{A}| |\mathcal{Y}|)^n)$ so this computation becomes intractable as the number of
bins $|\mathcal{Y}|$ or the number of steps $n$ grows.
     We are
  thus also interested in the performance of heuristic policies that are more
  easily computed, especially in the  regime where they can be compared to the
  optimal policy.
An heuristic policy we choose for this comparison is
the minimum posterior entropy
policy that minimizes the average entropy of the initial state conditional on
the next $g$ observations, as detailed in App.~\ref{sec:supadaptive}. Here we pick $g = 2$.
To understand the effectiveness of using active permutation policies,
we would like to compare 
 these policies
 to the policy of
 applying no actions at all. 
   To demonstrate the gain in fidelity achieved by using timing information,
   we also compare to a discrimination method that
   uses only aggregate probability of collecting numbers of photons over
   a total detection period, a method we refer to as the histogram method.
We do not
bin the distributions of total photons when computing infidelities for the
histogram method.

      To summarize, the four methods we compare are as follows.
      \begin{equation}
        \begin{tabular}{|c|c|c|}
          \hline
        Name & Permutation policy & Model  \\
        \hline
        Histogram & None applied & Total counts\\
        No Perms & None applied & HMM \\
        Min Entropy & Minimize expected entropy of next two observations & HMM\\
        Exhaustive & Minimize infidelity  & HMM\\
        \hline
        \end{tabular}
          \label{tab:methods}
      \end{equation}
 The ``Model'' column indicates which model is used for the different methods. ``Total counts'' means that the model
    ignores timing information, instead using just the total number of
    photons observed. ``HMM'' means that the model used is an HMM with a binned
    output distribution. We infer the initial
    state using maximum likelihood in every method, which incorporates the corresponding
model and applied permutations.

\begin{figure}
  \centering
    \def\svgwidth{\columnwidth}
    \scalebox{1}{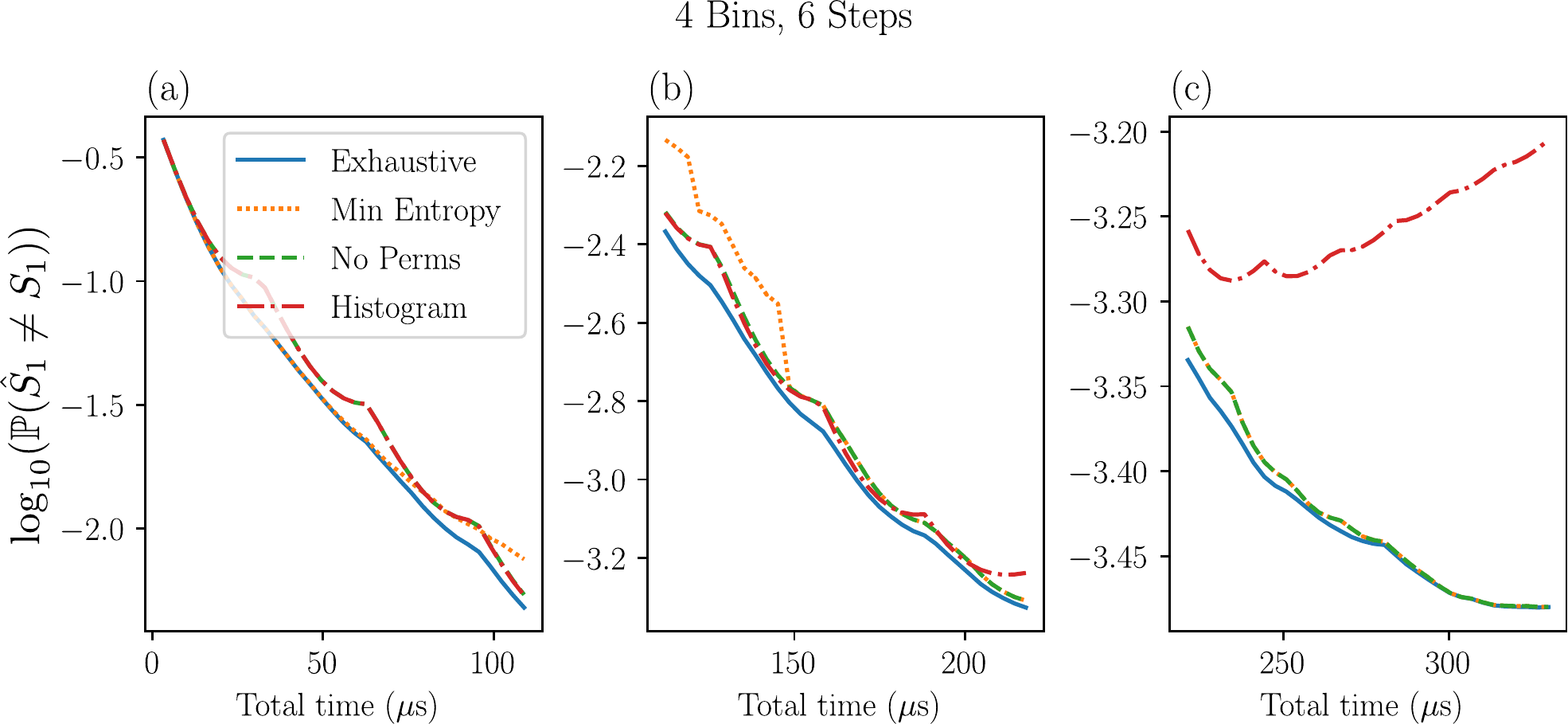}
  \caption{ A comparison of various measurement infidelities 
    (Eq.~\ref{eq:infidelityinterp}) under the uniform prior over the two states
  $(F, m_F) = (2, 2)$ and $(F, m_F) = (1, -1)$, for a range of collection
times. The infidelities are computed after binning the output distributions
into 4 bins
(except the Histogram method, which is not binned), for 6 steps of the HMM. Subfigures (a), (b), and (c) all show
the same infidelities plotted for different time ranges with re-scaled vertical
axes to highlight the relevant features. The dash-dotted red curve, labeled ``Histogram'', is the
infidelity when we do not apply permutations and  do not time-resolve the
  measurements into steps. This involves collecting photons for the time
  shown and determining the
  measurement outcome based on the total number collected. We note that in subfigure (c),
  Histogram curve begins to curve upwards. The dashed green curve is the
infidelity obtained by using an HMM, without applying any actions. In subfigure
(a), it agrees with the Histogram curve, but in subfigure (c), it outperforms
the Histogram curve. The solid
blue curve is the exhaustive search policy,
which is guaranteed to be optimal. It outperforms all other curves
in the figure, but has essentially no gain over the No Perms curve in subfigure
(c). The dotted orange curve is the minimum posterior entropy heuristic
policy, which performs well in subfigure (a), performs much worse
than all other curves in subfigure  (b), then recovers performance in subfigure
(c) end, coinciding with the No Perms and
Exhaustive curves there. The Min Entropy policy
was computed with a look ahead of
$g=2$ steps. }
  \label{fig:exact}
\end{figure}

We have plotted in Fig.~\ref{fig:exact}  the measurement infidelity
$\mathbb{P}(\hat{S}_1 \neq S_1)$ of
various methods of estimating the initial state, as a function of $n \Delta t$.
  For very small total times,  transitions are rare overall,  and therefore there is nothing to
  be gained by performing the time-resolved readout as we do in this work; it
  suffices to instead use the histogram method that performs state
  classification based solely on the aggregate 
  number of photons.
    As shorthand, we refer to the various curves in Fig.~\ref{fig:exact} by
    their legends. 

  Note that the regime of small times leads to large measurement error for
  any strategy chosen.
  At around $20 \mu$s, the Min Entropy and Exhaustive curves diverge from the other two.
  Here there is a gain due to applying the permutations, but the histogram
  method still coincides with the HMM method that does not apply permutations.
  Near $70 \mu$s, the Min Entropy curve diverges from the Exhaustive curve,
  indicating that our heuristic minimum posterior entropy policy underperforms
  the optimal one. At about $100 \mu$s, the minimum posterior entropy
  heuristic underperforms even the histogram method.
  At around $145 \mu$s, the Min Entropy curve rejoins the No Perms curve.
  Near $155 \mu$s, the No Perms and Histogram curves slightly diverge, indicating
  that our method of binning the outcomes into 4 bins causes us to incur a loss
  of distinguishability. Close to $200 \mu$s, the Min Entropy and No Perms
  curves diverge from the Histogram curve, indicating that it is worth using a time-resolved
  method for inference in this regime. Finally, beyond this point, the
  non-histogram methods begin to coincide. This indicates that there is nothing
further to be gained from active policies for large $\Delta t$.

  We see that the Min-entropy heuristic policy performs unpredictably in the
  different regimes. This is due to the fact that it does not incorporate
  information from all possible future paths. If the policy
  diverges from the optimal policy at any step, the resulting states that are
  realized are different, and the likelihoods of different possible $y^n$
  change dramatically. The best method to check for the performance of a heuristic policy is to directly calculate the infidelity, as we do in this paper.

\section{Example: Three-state model}
\label{sec:threestates}

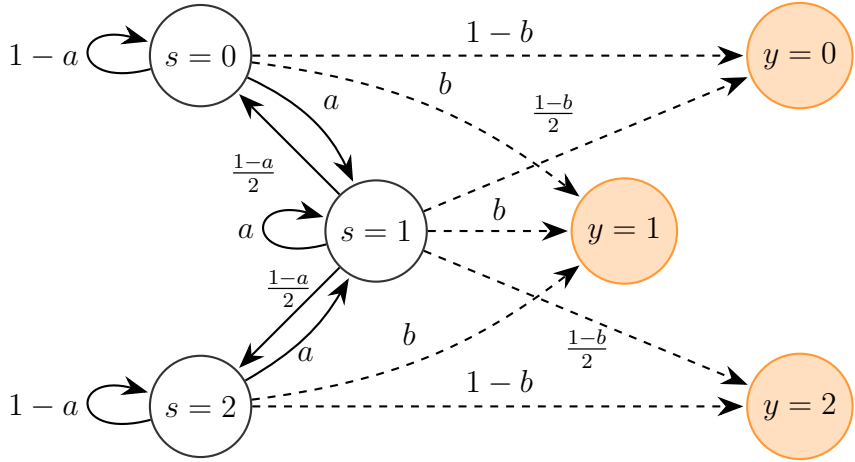
\begin{figure}[h]
\tikzstyle{state}=[circle, thick, minimum size=1.2cm, draw=black!80]
\tikzstyle{measurement}=[circle, thick, minimum size=1.2cm, draw=orange!80, fill=orange!25]
  \begin{tikzpicture}[->,>={Stealth[scale=1.5]},shorten >=1pt,auto,node
  distance=3.3cm, thick]
    \node[state]         (A)                    {$s=0$};
  \node[state]         (B) [below right of=A] {$s=1$};
  \node[state]         (C) [below left of=B]  {$s=2$};
  \node[measurement]        (E) [right of=B]       {$y=1$};
  \node[measurement]        (D) [above right of=E] {$y=0$};
  \node[measurement]        (F) [below right of=E] {$y=2$};

  \path (A) edge [bend left=20] node {$a$} (B)
            edge [loop left] node {$1-a$} (B)
            edge [dashed] node {$1-b$} (D)
            edge [bend left=15, dashed] node {$b$} (E)
        (B) edge [loop left] node {$a$} (B)
        edge  []  node {$\frac{1-a}{2}$} (A)
        edge  [above]    node {$\frac{1-a}{2}$} (C)
        edge [dashed] node {$b$} (E)
        edge [dashed] node {$\frac{1-b}{2}$} (D)
            edge [below, dashed] node {$\frac{1-b}{2}$} (F)
            (C) edge [below, bend right=15]   node {$a$} (B)
            edge [loop left] node {$1-a$} (C)
            edge [bend right=15, dashed]   node {$b$} (E)
            edge [dashed] node {$1-b$} (F);
\end{tikzpicture}
\caption{A toy example of a Hidden Markov Model on three states.
States are depicted by empty circles and outputs by colored circles. Arrows between
states indicate allowed transitions and are labeled with the probabilities in the transition
matrix. Arrows from states to outputs indicate possible outputs and are labeled
with the output probabilities.
We take the initial state distribution to be uniform on the three states. 
}
\label{fig:threestates}
\end{figure}

As a toy example, we consider a simple three-state model
that illuminates the advantages that can be gained 
 with permutation policies. 
The model is shown in
  Fig.~\ref{fig:threestates}.  The model has three states $\cS =
  \{0,1,2\}$ and three possible outputs $\cY=\{0,1,2\}$. 

  To keep the
  number of HMM parameters  low, it is symmetric under interchange of the
  labels $0$ and $2$ of both states and outputs. 
The transition rates and probability of outcomes shown in Fig.~\ref{fig:threestates} depend on two
  parameters, $a$ and $b$, which are intended to be small compared to 1. In this case, the state $s=1$ has a short lifetime in terms of the number of steps and quickly
  transitions  to $s=0$ or $s=2$ with equal probability, at a much larger rate
  than $0$ and $2$ transitioning to $1$.  States $0$
  and $2$ do not transition directly between each other and have  small probabilities of transitioning back
  to $s=1$  compared to the rates out of $s=1$. The output $y=0$ is only possible for
  states $s=0$ and $s=1$, while the output $y=2$ is only possible for
  $s=1$ and $s=2$. The output $y=1$ is equally likely for all states,
  and has low probability. Thus the output $y=0$ excludes $s=2$ and
  $y=2$ excludes $s=0$, while $y=1$ has no information about which
  state produced it.  We assume that the initial state distribution is
  uniform on the three states and aim to infer the initial state from
  the outputs produced after up to six steps.

\begin{figure}[h]
  \centering
    \def\svgwidth{\columnwidth}
    \scalebox{1}{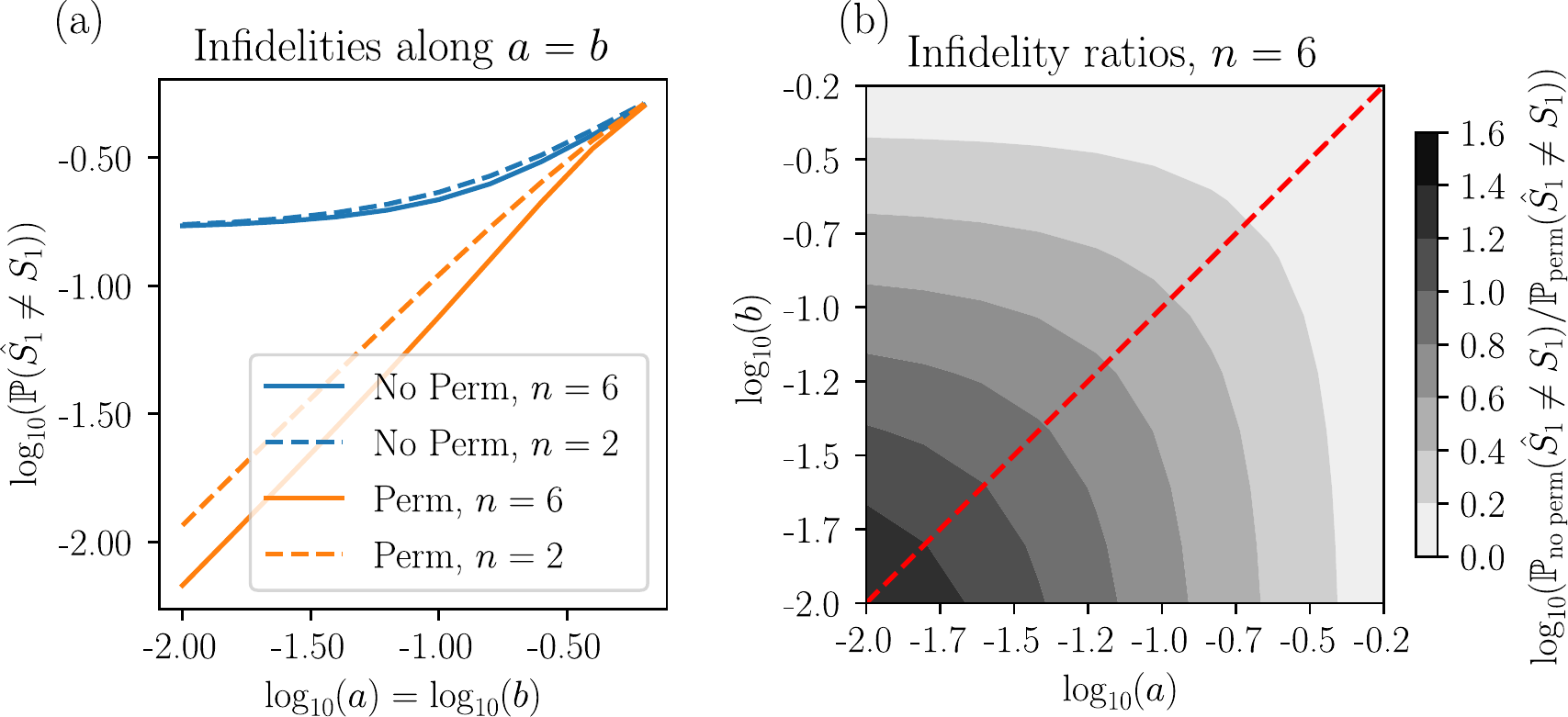}
    \caption{(a): The measurement infidelity (Eq.~\ref{eq:infidelityinterp}) of the measurement at 2 and
    6 steps, for both the maximum likelihood inference strategy without permutations and that with
    permutations. Although not demonstrated in this figure, the measurement infidelity in both cases stabilizes after
    6 steps, and we therefore do not show larger numbers of steps.
    The infidelity is calculated under the uniform prior,
      $\mathbb{P}(s_1) = 1/3$. The horizontal axis is a cut in the parameter space given by
        the line $a = b$, or in other words along the dashed line in (b).
        (b): The logarithm of the ratio of the measurement infidelity of the
        maximum likelihood
        estimator without permutations to that with permutations, for a range
        of parameters $a, b$. This metric is a measure of fidelity gained by implementing
        permutations over doing the same task without applying permutations.
        The gain is substantial for small $a$ and $b$. 
        \label{fig:threestategain}}
\end{figure}

We computed the measurement fidelities both for the  policy without permutations and
the  policy with the permutations selected according to the optimal policy
at $n=6$   steps
for a range of parameters $a$ and $b$. Here the permutations we allow are the transpositions between
any two states. We choose to stop after $n=6$ steps
because we found that the measurement infidelity changes very little
after reaching 6 steps, for the range of parameters shown
in Fig.~\ref{fig:threestategain}.

The gain in distinguishability is shown in
Fig.~\ref{fig:threestategain}~(b), which shows the ratio of the measurement infidelities without and with the permutation policy.
 We see that for small $a$ and small $b$ this ratio is large, demonstrating the relative improvement achieved. 
We can reason through this gain in distinguishability as follows.
Consider the case where the first observation is $y_{1}=0$. This
excludes the possibility that $s_{1}=2$, but the posterior
distribution has support on both $s_{0}$
($\Prob(S_{1}=0|y_{1}=0)=2/3$) and $s_{1}$
($\Prob(S_{1}=1|y_{1}=0)=1/3$).  If $s_{1}=1$ and no action is
taken, the state is likely to transition to $s_{2}=0$ or $s_{2}=2$,
with equal probability.  The transition to $s_{2}=0$ results in loss
of memory that $s_{1}=1$.  This is prevented by swapping the states
$1$ and $2$. In this case the likely next output is $y_{2}=2$ which
excludes the possibility that $s_{1}=0$. On the other hand, if
$s_{1}=0$, then the next outcome is likely again $y_{2}=0$, and we can
exclude the possibility that $s_{1}=1$.  A similar improvement in
distinguishing the initial states is obtained when $y_{1}= 2$, in
which case it helps to swap the states $0$ and $1$.
Thus, an adaptive
choice of permutation based on the first output improves the
measurement fidelity for this model.
Furthermore, because the swap required to achieve the improvement
depends on the first outcome, any outcome-independent choice of action performs less
well for two steps.

The pattern leading to the adaptive improvement of the three state
model generalizes.  Suppose that the outputs so far significantly
narrow the likely initial states to a subset of states, with memory of
the initial state still present but the states within the subset not
easily distinguished by future outputs.  Whenever this is the case,
one can gain an advantage by moving some of the states of this subset to
another set of more distinguishable states. The three-state model
provides a situation where  an adaptive  permutation policy is 
  strictly better than every non-adaptive such policy.

\section{Discussion}

\label{sec:discussion}

We have investigated policies for applying permutations between steps of
a repetitive measurement.
In our treatment of the measurement process, we have assumed that the
HMM parameters are known. In atomic systems, it is often possible to
determine these parameters from the physical constants associated with
the atomic levels, along with measured parameters such as Rabi rates. In
quantum systems with 
superconducting qubits~\cite{doi:10.1146/annurev-conmatphys-031119-050605} or
electrically-defined quantum dots~\cite{medford2013quantum},
the physics is less constrained, and it may be difficult to infer the
HMM from the measured or calibrated parameters. Instead, it is
possible to learn the HMM by observing the measurement process for
many steps.  Many tools are available for inferring the transition and
output matrices from such observations~\cite{Ephraim2002}. 
We recommend investigating use of these tools, while taking advantage
of known physical constraints and measured or calibrated parameters,
with the aim of improving the modeling of quantum measurement
processes.
  As mentioned in Sec.~\ref{sec:expimp}, when implementing permutations in systems
  with shorter lifetimes, the transition matrices and output matrices become
  directly dependent on which permutation is applied. Tools used for inference
  of HMMs can be adapted to the problem of inference of these action
  dependent models.

In specifying and applying permutation policies for improving measurement
fidelity, we have assumed that the permutations are applied perfectly.
As noted in \cite{Woelk2015,Hemmerling2012}, this assumption may not be
realistic.
The model can be readily adapted to take account of errors in applying
the permutations. Changing the transition matrix associated
with the chosen permutation to a general Markov process poses no difficulty
and makes it possible to account for known errors or noise in applying
the chosen permutation.

  In the absence of transitions, and where any permutation
  is an allowed action, optimizing the measurement infidelity is an
  instance of an active sequential hypothesis testing
  problem~\cite{10.1214/aoms/1177706205}. Computing the optimal policy
  for this type of problem is substantially simpler. The complexity is
  dominated by the number of possible posterior distributions on the
  input states given the outputs. This number  grows as a polynomial with the number
  of steps, although its degree may be high depending on the number of allowed actions.  For these problems, there are guarantees
  that the greedy policy used in this work has close to optimal mutual
  information between the initial state and the
  data~\cite{chen2015sequential}. One may also be interested in
minimizing the number of measurement rounds, instead of fixing it as we
have done in this work. One way to study this problem is to introduce
a ``discount factor'' that exponentially suppresses rewards that are
reached after many measurement rounds. One can then obtain upper and
lower bounds on the minimal discounted measurement
infidelity~\cite{kartik2019active}. A potential line of future work is to extend
these results to the case of non-trivial transition matrices.

In general, a quantum measurement process can
be described by a quantum instrument.  A quantum instrument has
classical outputs and output-conditional side-effects on the quantum system
being observed. One way to specify the instrument is as the
composition of a minimally disturbing general quantum measurement,
given as a positive operator-valued measure (POVM), followed by
output-conditional quantum operations. See Ref.~\cite{wilde2011classical} for technical
details. 
Because the processes of interest in this work were totally dephasing,
it was sufficient to use a classical description, but it is also of
interest~\cite{barry2014quantum} to study optimal policies for
measurement processes whose instruments are not totally dephasing.

In many situations, including those involving $^9\text{Be}^+$
  qubits, the physically implemented measurements involve
  continuous monitoring, and events such as detection times of photons
  are recorded. 
  The formalism used in our work requires discretizing
  time to allow us to model measurement of $^9\text{Be}^+$ as a discrete time Markov process. To take
  full advantage of the measurement process requires modeling by a
  continuous-time process\cite{combesRapidMeasurementPurification2010},\cite{PhysRevLett.100.160503}. 
This problem can be formalized with the Hamilton-Jacobi-Bellman
equation~\cite{kirk2004optimal},\cite{alt2020pomdps}.

The $^9\text{Be}^+$ example has the property that the graph of allowable
transitions is directed and acyclic. An HMM with this property is called
a left-right HMM, and simpler algorithms exist for computing likelihoods in
this degenerate case~\cite{Ephraim2002}.  While we did not take advantage of this structure in this
work, it is likely that using this structure would lead to simpler algorithms
for computing the optimal policy.

In this study, we focused on optimizing measurement fidelity given the
number of steps of the HMM measurement model that are observed.  The
HMMs relevant for atomic measurement eventually lose memory of the
initial state, so observing for more steps yields rapidly diminishing
fidelity gains. In many applications, it is desirable to minimize the
average time required to complete a measurement, which implies a
trade-off between measurement time and measurement fidelity.  Examples
where measurement time matters are for applications involving
feed-forward such as quantum error
correction, and in characterization
experiments dominated by measurement time.  The techniques discussed
here can be used to explore the measurement-fidelity measurement-time
trade-off. However, there is a way in which one can reduce average
measurement time, possibly without losing measurement fidelity.  In
particular, it is possible to terminate the measurement early if the
outputs so far indicate a particular initial state sufficiently
strongly. Such a scheme was introduced in Ref.~\cite{Myerson2008} and
related approaches are in current use~\cite{todaro2020}. HMMs can be used to improve these
schemes' time and fidelity performance beyond what can be achieved using
likelihood ratio tests computed from models with independent and identical
outputs. 
One can incorporate the cost of an additional measurement explicitly in the cost
function to minimize, and again consider the advantage that could be gained by
implementing an adaptive strategy.
In the case of a transition matrix equal to the identity, this problem has been
studied in~\cite{naghshvar2013active}. There, the authors take as their cost
function the expected number of measurements  plus the expected measurement infidelity, with a variable weighting between the two terms. Upper and lower bounds
on the optimal cost are then obtained.  It would be interesting to extend these results to the scenarios
with general transition matrices considered here.

\section{Conclusion}
\label{sec:conclusion}

We have investigated the use of adaptively chosen actions to
  improve measurement fidelity in quantum measurements that are
  realized as sequential observations with complete decoherence. We
  considered two examples, one motivated by $^9\text{Be}^+$ ion qubits, the other
    a three-state toy example. We focused on
  actions consisting of permuting the states of the
  system. 

  Our study of $^9\text{Be}^+$ measurements
  indicates parameter regimes where an improvement is achieved and
  suggests future work to take advantage of adaptive permutation
  policies.  We discussed a number of paths forward and open problems,
  such as that of optimizing the trade-off between measurement times
  and measurement fidelity, finding better policies, and extensions to
  continuous-time measurement processes.

\appendix

\section{Outline of Appendix}
In the appendices below, we explain the mathematical formalism involved in the
calculations used in the main text. In App.~\ref{sec:repetitive} we introduce repetitive quantum
  measurements. We discuss why, in many experiments, it is possible to
  describe such measurements with an effective classical model, namely an 
  HMM. In App.~\ref{sec:formalismexample} we discuss a few subtleties involved in discretizing
  continuous-time dynamics so that it can be modeled as an HMM, illustrating our solution in
  the context of the $^9\text{Be}^+$ example.  In
  Sect.~\ref{sec:initial} we describe the problem of inferring the
  initial state from a sequence of measurements and its solution by
  the maximum likelihood estimate.  The method of using adaptively
  chosen actions during repetitive measurements is defined in
App.~\ref{sec:supadaptive}.  Therein, we discuss how to compute the optimal
policy using the Bellman Equation,
and present our heuristic policy that chooses actions based on minimizing the entropy of the
initial state. We also sketch how to reduce the problem of
  computing an optimal policy to a POMDP in App.~\ref{sec:supadaptive}. Finally, our implementation and simulation is
  outlined in App.~\ref{sec:impsim}.

\section{Repetitive measurement models}
\label{sec:repetitive}

Here we discuss the formalism of Hidden Markov Models (HMMs), and show how they
describe
repeated measurements of quantum systems.
As discussed in the main text, for a quantum system under interrogation by
a measurement process, such as an atom under fluorescence detection, the state
of the system can be effectively dephased, and thus has a classical description.
The initial quantum
state is a probability distribution over the levels $\mathcal{S}$ (assumed
to be nondegenerate) that constitute
the measurement basis elements, and that the output has a probability
distribution conditional on the current level and previous level, and the next level
has a probability distribution conditional on the current level.
This is immediately in the form of an HMM step, except that
the current outcome may depend on the previous level as well
as the current level.
The dependence on the previous level can be accounted for by expanding the HMM state space to include memory
of the previous outcome, as explained below. For  atoms with cycling
transitions, the transitions between levels result from non-ideal cycling.
Having reviewed the reduction from a quantum model to a classical
  stochastic one, we now use the term ``state'' to refer to a classical level.

We use the convention that an upper case
  variable $X$ refers to a random variable (RV), while its corresponding lower
  case variable refers to a particular instantiation of the RV. We use $\Prob$ to denote ``probability of'', whose argument is an
  event, so that the expression $\Prob(X = x)$ refers to the probability that
  the RV $X$ has a particular value $x$. Because we are interested in
  stochastic processes, we also use the notation for sequences of RVs as in the
  main text. A variable $S_i$ is the RV for the stochastic process $S$ at
  the step $i$, while the variable $S^i$ indicates the subsequence of the first
  $i$ steps of the process $S$, $S^i = S_1, \ldots, S_i$. We adopt similar
  conventions for particular sequences of values, so that $s^i = s_1, \ldots,
  s_i$.

We now introduce notation for HMMs. An HMM has state space $\cS$, output space
$\cY$, transition matrix $A$, and output matrix $B$.
We use the
notation $A(s'|s)$ for the transition probability to next state $s'$
given current state $s$, and $B(y|s)$ for the probability of output
$y$ given current state $s$. We denote the state and observation at step $i$ as $s_i$ and $y_i$, respectively.
Let $\nu$ be the initial state distribution
defined by $\nu(s_{1}) = \Prob(s_{1})$, where the expression $s_{1}$ abbreviates the
event that $S_{1}=s_{1}$. 
The probabilities of the
state and output sequences are determined by unraveling the
transitions according to
\begin{align}
  \Prob(s^{n}) &= A(s_{n}|s_{n-1})\Prob(s^{n-1}) = \nu(s_{1})\prod_{i=2}^{n}A(s_{i}|s_{i-1}),\nonumber
  \\
  \Prob(y^{n}|s^{n})&= \prod_{i=1}^{n}B(y_{i}|s_{i}).
  \label{eq:HMM}
\end{align}

\section{Example: Fluorescence detection of $^9\text{Be}^+$
hyperfine qubit}
\label{sec:formalismexample}

  To describe the measurement dynamics of $^9\text{Be}^+$ measurement as an
  HMM, we address the following issues. While dynamics of physical systems
  usually take place continuously in time, an HMM is a discrete-time model, and
  therefore the physical dynamics need to be discretized. Second, the distributions
  of outcomes can contain more information than is useful, which can complicate computations.
  We thus simplify the outcome distributions by a binning procedure.
   Third, as mentioned
  in the introduction, the probability of transitioning to a different state
  sometimes depends on the measurement outcome.
  This is accounted for by expanding the state space to include the current
  measurement outcome, allowing the transition matrix $A$ to depend on both the physical state and the
  outcome.

\paragraph*{Reduction from continuous to discrete time}
\label{sec:ctstimereduction}
An HMM description
of a fluorescence measurement can be derived from the continuous time
Markov process modeling the stochastic dynamics of the levels
and  the detection of fluorescence from the cycling level while
driving the cycling transition. 

Transitions between levels are described by the transition-rate matrix $Q$.
The off-diagonal entries $Q_{ss'}$ of $Q$ are the non-negative transition rates,
and the diagonal entries $Q_{s's'}=-\sum_{s\ne s'} Q_{ss'}$ are the total rates of departure
from level $s'$ to other levels.
The transition-rate matrix can be
integrated to obtain the probabilities $R(s|s')$ of starting in level $s'$ and
ending in level $s$ for a measurement step of period $\Delta t$. Then $R(s|s')
= (e^{Q\Delta t})_{s,s'}.$ The photon emission
rate for level $s$ is $E_s \geq 0$. For simplicity, we make the approximation
that at most one transition occurs in any given step. If the ion is in
level $s_{i-1}$ at the beginning of step $i$, and in level $s_i$ at the end of the
step, and transitions at a particular time $t$, the distribution of
collected photons is Poissonian with mean 
determined by $E_{s_i}, E_{s_{i-1}},$ and $t$. The distribution $J(o|s, s')$ of the number $o$ of
collected photons given the system starts in level $s'$ and ends in
level $s$
is then the mixture of these Poisson distributions~\cite{Langer2006}. We can compute the probability of the system
starting the step in the level $s'$, observing $o$ photons during the
step  and  ending in level $s$ as $U\left( s,
o \middle| s' \right) = J(o|s, s') R(s|s')$.

  The output distributions are supported on all nonnegative integers, but
  collecting a large number of photons in a single step is very unlikely.
  We thus restrict the output space to be the set $\mathcal{Y} = \left\{ 0, \ldots,
    n_{\text{max}}-1, n_{\text{max}}
  \right\}$, where we obtain outcome $n_{\text{max}}$ if at least
  $n_{\text{max}}$ photons were collected during the step.
We chose $n_{\text{max}}
  = 15$ so that the probability of collecting $n_{\text{max}}$ or more photons from the bright state in a time step of duration $\Delta t = 53.9 \mu$s is less than $10^{-3}$.

\paragraph*{Outcome dependence}
\label{sec:outcomedependence}
In the model described in the previous paragraph, the output
  depends on the previous and current state. In
  order to describe such models, we introduce notation for HMM state
  spaces that expand the physically relevant state space. Let $\cS$ be
  the set of physical states. The state space $\cT$ of the
  expanded HMM is related to the physical states by a map $\alpha:\cT\rar
  \cS$. In the $^9\text{Be}^+$ example, we let 
  $\cS$ be the set of levels, $\cS=\{(F, m_F)|F \in
  \{1, 2\}, m_F \in \{-F, \ldots, F\}\}$.
  We let $\cT$ consist of pairs $(s,o)$ of
  physical states $s\in \cS$ and outputs $o\in\cY$, and we define
  $\alpha((s,o)) = s$. Here, $s$ represents the current physical
  state and the $o$ is the output observed during the step resulting
  in state $s$.  For this example, it is also useful to define
the map $\rho:{\cal T} \rar {\cal Y}, \rho((s, o)) = o$. We
then define the transition and
output matrices as follows.
The transition probability from
previous state $s'$, with previously recorded observation $o'$, to the
current state $s$ with observation $o$ recorded in transitioning to
$s$ is
\begin{align}
  A\left( s, o \middle| s', o' \right) = U\left( s, o \middle| s' \right).
  \label{eq:AU}
\end{align}
Because the transition matrix now captures both the state transition and the probability of a state emitting an outcome,
the outcome process matrix on the expanded state is deterministic, so that $B(o|l) = \delta_{o, \rho(l)}$ is the new outcome process matrix.
Thus, the outcome matrix merely describes the fact that the physical state is unobservable.
  For the initial state distribution, we must pick a convention for the mapping
from the physical initial state distribution to that for the HMM. Let $S_1$ be
the random variable describing the initial physical state. 
For the fluorescence example  we take the
  distribution over initial HMM states to be $\nu(s, o) = \Prob\left( S_1 = s
  \right)\delta_{o, 0}$.

  \paragraph*{Binning}
  \label{sec:binning}
For our purposes, the outcome distributions contain more information than
necessary, and the computations grow rapidly in complexity with the number of
outcomes.
Given an HMM $M$ with an output space $\mathcal{Y}$, we thus seek a related model
that has fewer possible outputs.
We accomplish this by partitioning the outcome space into bins.

A partition $\mu = \{ b_i \}$ of
$\mathcal{Y}$ with $n_b$ bins is a set of $n_b$ nonempty  subsets of the set
$\mathcal{Y}$, such that
the sets $b_i$ are pairwise disjoint, $i\neq j \imp b_i \cap b_j = \emptyset$,
and cover the set
$\mathcal{Y}$, $\cup_i b_i = \mathcal{Y}$. An element $b_i \in \mu$ is called
a bin.
The set of bins $\mu$ becomes the output space of the binned model $M_\mu$.
  By keeping only the information of which bin the outcome lies in, the
  resulting distribution is again an HMM.

For the $^9\text{Be}^+$ example, the binned model has initial state distribution $\nu_\mu$, transition matrix
$A_\mu$, and output matrix $B_\mu$, defined by
\begin{align}
  \nu_\mu(s, b_i) &=  \sum_{y \in b_i}\nu(s, y)\\
  A_\mu(s, b_i|s', b_j) &= \sum_{y \in b_i}U(s, y|s')\\
  B_\mu(b_i|s) &= \sum_{y \in b_i}B(y|s).
  \label{eq:condbinning}
\end{align}
Note that the bins cannot depend on the state.

To determine what bins to use, it is useful to have a
quantitative cost function to apply to the bins. 
  Specifically, if $f$ is a function that takes an HMM $M_\mu$ and gives a cost
  associated with that model, we wish to choose the binning $\mu^*_{n_b}$ with
  $n_b$ bins that solves the minimization problem
  \begin{equation*}
\begin{aligned}
& \underset{\mu}{\text{argmin}}
& & f(M_\mu) \\
& \text{subject to}
& & \mu \vdash \mathcal{Y} \\
&&& |\mu| = n_b.
\end{aligned}
\end{equation*}
  where the notation $\mu \vdash \mathcal{Y}$ denotes that $\mu$ is a partition
  of $\mathcal{Y}$. 
We use the infidelity (Eq.~\ref{eq:infidelityinterp}) of the model induced by the
binning as our cost function $f$ to determine the optimal binning when we compute policies in
Sect.~\ref{sec:be9}, but in general different heuristic costs can be
used~\cite{keith2018joint}.

\section{Initial state inference}
\label{sec:initial}

Given the measurement process described in App.~\ref{sec:repetitive}, our goal is to infer, after some number of
  observations, the initial physical state of the system.
  So consider now a generic HMM with state space $\cT$,  physical state space
  $\cS$  identified by $\alpha:\cT\rar \cS$, initial physical state space $\cL$,
  transition matrix $A$ and output matrix $B$.
  We assume that the initial state is uniformly distributed
  over $\cL$.
  Let $\hat S_{1} = \phi(Y^n)$ be the random variable that is the output of the initial-state estimation procedure $\phi$ determined by the observed outputs $Y^n$.
    Recall the definition of the measurement infidelity in Eq.~\ref{eq:infidelityinterp},
    $\Prob(S_1 \neq \hat{S}_1)$. We will sometimes also refer to the fidelity $F = \Prob(S_{1}= \hat
    S_{1}) = 1-\Prob(S_1 \neq \hat{S}_1)$.

The maximal
measurement fidelity is achieved by choosing this function to be the Bayesian, maximum a-posteriori
(MAP) estimate~\cite{murphy2012machine}.  The MAP estimate $\hat s_1$ is the physical state with the highest posterior probability
given the $n$ observed outcomes:
\begin{align}
   \hat s_1 &= \argmax_{s_1\in \cS} \Prob(s_1|y^{n}).
   \label{eq:mapestimator}
\end{align}
According to Bayes's rule, $\Prob(s_1|y^{n}) =
\Prob(y^{n}|s_1)\Prob(s_1)/\Prob(y^{n})$. Because the denominator is
independent of $s_1$   and the prior distribution is uniform,
The MAP estimate is the same as the maximum likelihood
(ML) estimate~\cite{shao2003mathematical}:
\begin{align}
   \hat s_1 &= \argmax_{s_1\in \cS} \Prob(y^{n}|s_1).
   \label{eq:physicalML}
\end{align}
This estimate can be computed step-by-step  by keeping track of
the  list of values $\left(\Prob(y^{k},
  l_{k}|s_1)\right)_{l_{k}\in\cT, s_1\in\cL}$ for $k=1$ to
$k=n$, where the list is updated by applying the recursive expression

\begin{align}
  \Prob(y^{k}, l_{k}|s_1)
    &= \sum_{l_{k-1}\in \cT}
            \Prob(y^{k-1},l_{k-1}|s_1)
            \Prob(y_{k},l_{k}|y^{k-1},l_{k-1},s_1)\nonumber\\
    &= \sum_{l_{k-1}\in \cT}
            \Prob(y^{k-1},l_{k-1}|s_1) \Prob(y_{k},l_{k}|l_{k-1})\nonumber\\
    &=\sum_{l_{k-1}\in \cT}
            \Prob(y^{k-1},l_{k-1}|s_1) B(y_{k}|l_{k})A(l_{k}|l_{k-1}),
     \label{eq:update}
\end{align}
which takes advantage of the HMM conditional independence properties.
The values are initialized with 
\begin{align}
  \Prob(y^{1},l_{1}|s_1)
  &= \left.B(y^{1}|l_{1})\mathbb{I}[s_{1}=\alpha(l_{1})]\nu(l_1)\middle/\sum_{l'_{1}: s_{1}=\alpha(l_{1}')}\nu(l_1')\right..
\end{align}

From the final values $\Prob(y^{n},l_{n}|s_1)$, the MAP estimate is
obtained according to
\begin{align}
   \hat s_{1} &= \argmax_{s_1 \in \cL}\sum_{l_{n}} \Prob(y^{n}, l_{n}|s_1).
\end{align}
By our construction of the expanded HMM, this is equal to the expression in Eq.~(\ref{eq:physicalML}).  The measurement fidelity $F$ can be computed exactly if the number of
possible outcome sequences $y^{n}$ is not too large, or by
empirically sampling the HMM output sequences and computing the Bayesian posterior probabilities
for each sample.

\section{Adaptive measurement policies}
\label{sec:supadaptive}

\paragraph*{Modification of HMM}
\label{sec:adaptive}

To accommodate actions that can be taken between steps of an HMM we
introduce a set of possible actions $\cA$, where each action $a$ in $\cA$ is a
state-transforming process. In general, $a$ can be a stochastic process,
where the probability that $a$ results in state $s'$ given that the
current state is $s$ is denoted by $a(s'|s)$. The process resulting
from modifying the HMM by applying action $a_{k}$ after observing
$y_{k}$ in the $k$\textsuperscript{th} step can be thought of as an HMM with
step-dependent transition probabilities, where the transition matrix
$A$ at step $k$ is replaced by the composition of $A$ with the action
$a_{k}$. The modified transition probabilities are then
\begin{align}
  A_{k}(s_{k+1}|s_{k}) &= \sum_{s}A(s_{k+1}|s)a_{k}(s|s_{k}).
\end{align}
The probability of a state sequence is accordingly given by
\begin{align}
  \Prob(s^{n}) &= \nu(s_{1})\prod_{i=2}^{n}A_{i-1}(s_{i}|s_{i-1}).
\end{align}
For computing the final estimate, it suffices to modify the expression
for $\Prob(y^{k},s_{k+1}|l_1)$ in Eq.~\ref{eq:update} by replacing
$A(s_{k}|s_{k-1})$ with $A_{k-1}(s_{k}|s_{k-1})$, which depends on the
policy's choice of action at step $k-1$.

The action at step $k$ is chosen based on the observations so far
given by $y^{k}$.  A policy  is a specific strategy for choosing
the action.     Policies are chosen to maximize a reward. If
the reward can be expressed as a sum of rewards at each step, this
fits the framework of partially observed Markov decision processes
(POMDPs)~\cite{Astrom65}. For our application, policies are chosen to
maximize the measurement fidelity, which is expressed in
terms of a decision made after the last step.  It is possible to change
the model to an equivalent one fitting the POMDP formalism, as discussed below.

We consider actions that permute states, so $\cA$ is a set of state
permutations including the identity permutation.  We denote such
permutations by $\sigma$ and write $\sigma(s)$ for the state resulting
from applying $\sigma$ to state $s$.  
The process is schematically shown in
Fig.~\ref{fig:phmm}. Note that there are $n-1$ actions for a sequence of $n$
steps.

  In the case that the HMM state space $\mathcal{T}$ does not coincide with the
  physical state space $\mathcal{S}$, it is necessary to relate
  physical actions to actions on the HMM state space.
  For the $^9\text{Be}^+$ system, recall that the HMM states are given by pairs $(s,o)$ of
  physical states $s\in \cS$ and outputs $o\in\cY$, where
  $\cS$ is the set of levels, $\cS=\{(F, m_F)|F \in
  \{1, 2\}, m_F \in \{-F, \ldots, F\}\}$, and $\cY$ is the set of photon
    counts, $\cY=\{n|n=0,1,\ldots, n_{\text{max}}\}$.
  Given a permutation $\sigma$ that acts on the physical states $\mathcal{S}$, we
  induce a permutation $\tilde{\sigma}$ that acts on $\mathcal{T}$, by taking
  $\tilde{\sigma}((l,o)) = (\sigma(l), o)$. The permutations for the $^9\text{Be}^+$
  example are then $\left\{ \tilde{\sigma}|\sigma \in \left\{ \tau, \tau^{-1},
\epsilon \right\} \right\}$, where the permutation $\tau$ is defined in
Fig.~\ref{fig:be9levels}.

\paragraph*{Bellman Equation}
\label{sec:bellman}

  Our task now is to find a policy, consisting of permutations to apply,
  that maximizes the measurement fidelity.
  For sake of clarity, we assume in this appendix that the HMM state space
  coincides with the physical state space, but the ideas discussed here can be
   extended when this is not the case, by maximizing the
  probability of inferring the initial physical state instead of the initial
HMM state. In the following, we formally express the maximization problem for
computing the optimal policy, then introduce belief states, and show their
use in computing the optimal policy.

To maximize the measurement fidelity
over the choices of $\sigma^{n-1}$, we use the law
  of total expectation to compute that
  \begin{align}
    \max_{\sigma^{n-1}}&\,\Prob\left( \hat{S}_1 = S_1 | \sigma^{n-1} \right)
      = \nonumber\\ &\mathbb{E}_{Y_1}\left(\max_{\sigma_1}\mathbb{E}_{Y_2|\sigma^1}\left(
    \cdots\max_{\sigma_{n-2}}\mathbb{E}_{Y_{n-1}|\sigma^{n-2}}\left(\max_{\sigma_{n-1}}\mathbb{E}_{Y_n|\sigma^{n-1}}\left(\Prob\left(
    \hat{S}_1 = S_1|Y^n \sigma^{n-1}
        \right)  \right)\right) \cdots\right)\right)\label{eq:bellman}
  \end{align}
  This is called the Bellman Equation~\cite{bellman1957dynamic},~\cite{kochenderferWheelerWray2022}.
One way to compute Eq.~\ref{eq:bellman} is to compute all possible posterior
distributions of the initial state $S_1$, then to compute the expectation
of the reward $\Prob( \hat{S}_1 = S_1|Y^n \sigma^{n-1} )$ for all possible
sets of actions $\sigma^{n-1}$.
The posterior distributions are
computed iteratively. To describe the computation, we introduce the distributions
  \begin{align}
    \beta_k(s_1, s_k|y^k, \sigma^{k-1}) = \Prob(s_1, s_{k}|y^{k}, \sigma^{k-1})
    \label{eq:nonoffsetbelief}
  \end{align}
  and
  \begin{align}
    \eta_k(s_1, s_{k+1}|y^k, \sigma^k) = \Prob\left( s_1, s_{k+1} \middle|
    y^{k}, \sigma^{k} \right).
    \label{eq:offsetbelief}
  \end{align}
  The $\beta_k$ and $\eta_k$ are called belief states.
  The belief state $\beta_k$ describes the state of knowledge 
    of the current and initial HMM states
  immediately after
  observing outcome $y_k$, but before we apply the permutation $\sigma_k$.
  After applying $\sigma_k$ and allowing the system to transition, the state of
  knowledge is described by $\eta_k$.
    Because our goal is to infer the initial state of the system, note that it
    is necessary to keep track of the state of knowledge of the initial state.
    To predict how the dynamics will effect our estimate, it is also necessary
    to have an estimate of the current state.
    We therefore keep track of $s_k$ as well in the belief state.

  The belief state is iteratively updated according
  to the observations $y^n$ and permutations $\sigma^{n-1}$.
  At step $k$, we use the observation
  $y_k$ to update the belief state 
  $\eta_{k-1}$  to $\beta_k$ according to Bayes' rule
  \begin{align}
    \beta_{k}(s_1, s_{k}|y^{k}, \sigma^{k-1}) &= h_{\text{Bayes}}(\eta_{k-1}, y_k) = \frac{\eta_{k-1}(s_1, s_{k}|y^{k-1}, \sigma^{k-1})B(y_{k}|s_{k})}{
      \sum_{s_{k}}\eta_{k-1}(s_1, s_{k}|y^{k-1},
      \sigma^{k-1})B(y_{k}|s_{k})}\label{eq:bayesupdate}.
  \end{align}
  Next, we use the permutation $\sigma_k$ to update the belief state $\beta_k$ is updated to $\eta_k$ according the transition matrix
  of the HMM,
\begin{align}
  \eta_k(s_1, s_{k+1}|y^k, \sigma^{k}) &= h_{\text{Trans}}(\beta_k, \sigma_k) =  \sum_{s_{k}}\beta_k(s_1, s_{k}|y^{k},
  \sigma^{k-1}) A(s_{k+1}|\sigma_{k}(s_{k})).\label{eq:transitionupdate}
\end{align}
  In Eq.~\ref{eq:bayesupdate}, we used the fact that the current outcome
  depends only on the current state, and in Eq.~\ref{eq:transitionupdate}, we
  used the fact that the next state depends only on the current state, which
  are the two defining features of the HMM.

  The initial belief state
  $\eta_0$ is determined by the prior distribution,
  $\eta_0(s_1, s_1'|y^0, \sigma^0) = \nu(s_1)\delta_{s_1, s_1'}$.
Here, the index $s_1'$ is for the ``current'' state after having seen no data
$(y^0)$ and having applied no permutations $(\sigma^0)$. This index is updated
after each application of $h_{\text{Trans}}$, updating the step index each
time.  In contrast, the other index $s_1$ is the one that describes the initial
state, and continues to describe the initial state even after the various
updates using $h_{\text{Trans}}, h_{\text{Bayes}}$.

\begin{figure}[h]
  \centering
\tikzset{
  treenode/.style = {shape=rectangle, rounded corners,
                     draw, align=center,
                     top color=white, bottom color=blue!20},
  eta/.style     = {treenode, font=\large, bottom color=blue!30},
  beta/.style    = {treenode, font=\large, bottom color=red!30}
}
\tikzstyle{level 1}=[level distance=5cm, sibling distance=2.3cm]
\tikzstyle{level 2}=[level distance=8cm, sibling distance=1cm]
\begin{tikzpicture}
  [
    grow                    = right,
    edge from parent/.style = {draw, -latex},
    every node/.style       = {font=\footnotesize},
    sloped
  ]
  \node [eta] {$\eta_0$}
  child { node [beta] {$\beta_1 = h_{\text{Bayes}}(\eta_0, y_1=1)$}
  child { node [eta] {$\eta_1 = h_{\text{Trans}}(\beta_1, \sigma_1=b)$}
                edge from parent node [below, align=center]
                {\large$\sigma_1 = b$}
            }
            child { node [eta] {$\eta_1 = h_{\text{Trans}}(\beta_1, \sigma_1=a)$}
                edge from parent node [above, align=center]
                {\large$\sigma_1 = a$}
            }
      edge from parent node [below, align=right] {\large$y_1=1$\,\,\,\,\,\,} }
      child { node [beta] {$\beta_1 = h_{\text{Bayes}}(\eta_0, y_1=0)$}
      child { node [eta] {$\eta_1 = h_{\text{Trans}}(\beta_1, \sigma_1=b)$}
                edge from parent node [below, align=center]
                {\large$\sigma_1 = b$}
            }
            child { node [eta] {$\eta_1 = h_{\text{Trans}}(\beta_1, \sigma_1=a)$}
                edge from parent node [above, align=center]
                {\large$\sigma_1 = a$}
            }
        edge from parent node [above, align=left] {\large$y_1=0$\,\,\,\,\,\,} };
\end{tikzpicture}
  \caption{A depiction of the belief state tree. For illustration purposes, we
    consider the case that $\mathcal{Y} = \left\{ 0, 1 \right\}$, and
    $\mathcal{A} = \left\{ a, b \right\}$ both have two elements. To keep the
  diagram small, we depict only one step of the calculation. We traverse the
tree by using Bayes' rule to update an $\eta$ belief state, or using
an action to update a $\beta$ belief state.}
  \label{fig:belieftree}
\end{figure}
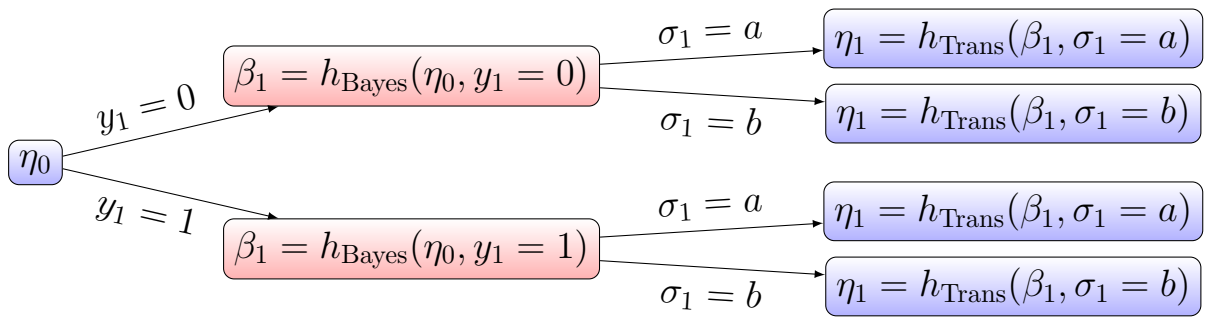

  The belief states are labelled by sequences of observations $y^n$ and
  permutations $\sigma^{n-1}$, and the set of all accessible belief states has the
  structure of a tree, see Fig.~\ref{fig:belieftree}. The children of a node $\eta_{k-1}$ of the tree are
  those $\beta_k$ that are obtained from a Bayes update as in
  Eq.~\ref{eq:bayesupdate}, where the different children are labelled by the
  possible values of $y_k$. Similarly, the children of a node $\beta_k$ are
  those $\eta_k$ that can be obtained from Eq.~\ref{eq:transitionupdate}, where
  the different children are labelled by the possible values of $\sigma_k$.

  Once all the belief states have been computed, the exhaustive search
  algorithm suggested by Eq.~\ref{eq:bellman} can be directly implemented.
  To perform this computation using the tree, we start at the leaves by
  computing the probability of correct inference 
  \begin{align}
  \Prob(S_1 = \hat{S}_1|y^n,
  \sigma^{n-1}) = \max_{s_1}\sum_{s_{n}}\beta_n(s_1, s_{n}|y^n\sigma^{n-1}).
    \label{eq:leafvalue}
  \end{align}
  Next,
  we compute the expectation
\begin{align}
  \mathbb{E}_{Y_n|\sigma^{n-1}}\left(\Prob(S_1 = \hat{S}_1|y^n,
  \sigma^{n-1})\right) = \sum_{y_n}\Prob(S_1 = \hat{S}_1|y^n, \sigma^{n-1})\sum_{s_n}B(y_n|s_n)\sum_{s_{1}}\eta_{n-1}(s_1, s_{n}|y^{n-1}\sigma^{n-1}).
  \label{eq:etaexpect}
\end{align}
This is the innermost expectation in Eq.~\ref{eq:bellman}.
We next compute the maximum over the permutation $\sigma_{n-1}$ directly.
We have thus computed the $(n-1)$th step of the optimal policy.
By then iterating expectation and maximization steps, we 
compute the optimal policy of Eq.~\ref{eq:bellman}.

\paragraph*{Minimum posterior entropy heuristic}
\label{sec:minent}

In the examples in Sect.~\ref{sec:threestates} and Sect.~\ref{sec:be9}, we compute the optimal policy using exhaustive search for a small 
set of actions and a small set of outcomes, but as the numbers of actions and outcomes increase, the exhaustive search algorithm becomes extremely expensive to compute.  This is because
 the work
required scales as $O( (|\mathcal{A}| |\mathcal{Y}|)^n)$, determined by the size of the tree. For more than a few steps, computing the optimal policy is currently infeasible. This motivates the use of heuristic policies.
 We here discuss an heuristic policy that performs well in some regimes, as
 a possible alternative to the optimal policy when it is not available.

Our heuristic policy is also a belief-state-based algorithm, but while
exhaustive search constructs a tree of height $2n+1$, we construct $O(|\mathcal{Y}|^n)$ smaller
trees of height $2g+1$ for a fixed $g$. 
To compute our heuristic, at each step we compute a measure of
concentration of the posterior probability distribution of the initial physical
state given the output of the next $g$ steps, and choose the
permutations that optimize the expectation of that measure.
This strategy is in a sense greedy  and its
performance depends on the measure of concentration used.  For this
study, we used the posterior entropy as a measure of concentration,
with less entropy indicating higher concentration.  We call this the
minimum posterior entropy heuristic.
This is similar to the policy
given in Ref.~\cite{Cassandraa}, except that there, the
authors minimize posterior entropy of the HMM state at the next step, not that of
 the initial state.  Note that measurement fidelity is
determined by the maximum probability of the posterior. Maximum
posterior probability is also a measure of concentration but is
insensitive to probabilities other than the maximum one, which may
result in blind spots  for a greedy algorithm using maximum
posterior probability.  
We distinguish here between the policy that involves maximizing the maximum probability of the
  posterior at $n$ steps and that at $g$ steps. The former is the optimal
  policy, since at $n$ steps the maximum probability of the posterior is the
  measurement fidelity, while at $g$ steps it is merely an (uncontrolled)
  approximation of the fidelity.

  As in App.~\ref{sec:supadaptive}, we assume in the following discussion that the HMM state space coincides with
the physical state space, but again the calculations can be  adapted to
a case where they differ, by changing the cost function to be the posterior entropy
of the physical initial state.
  For choosing the permutation to be applied at step $k$ after having observed
  $y^k$ and having applied actions $\sigma^{k-1}$, we compute
  all possible belief states after $g$ more steps, given by
  \begin{align}
    \Prob\left( s_1, s_{k+g} \middle| y^{k}, \sigma^{k-1}, y_{k+1}^{k+g},
    \sigma_{k}^{k+g-1} \right),
    \label{eq:smalltree}
  \end{align}
  for all possible choices of $y_{k+1}^{k+g}, \sigma_{k}^{k+g-1}$,
  where we use the abbreviation $y_{k+1}^{k+g}$ to mean the subsequence
  $y_{k+1}, \ldots y_{k+g}$, and similarly for $\sigma_{k}^{k+g-1}$. 
  We can then compute the associated initial state distributions,
  \begin{align}
    \Prob\left( s_1 \middle| y^{k}, \sigma^{k-1}, y_{k+1}^{k+g},
    \sigma_{k}^{k+g-1} \right) = 
    \sum_{s_{k+g}}\Prob\left( s_1, s_{k+g} \middle| y^{k}, \sigma^{k-1}, y_{k+1}^{k+g},
    \sigma_{k}^{k+g-1} \right),
  \end{align}
  and the corresponding entropy
  \begin{align}
    H(S_1|y^{k+g}, \sigma^{k+g-1}) = -\sum_{s_1}\Prob\left( s_1 \middle|
    y^{k+g}, \sigma^{k+g-1}
    \right)\log\left( \Prob\left( s_1 \middle| y^{k+g}, \sigma^{k+g-1} \right)
    \right).
    \label{eq:initialentropy}
  \end{align}
  Viewing this entropy as a cost function, we can then choose the permutations
  $\sigma_{k}^{k+g-1}$ that minimize this cost, and apply the permutation
  $\sigma_{k}$. 

  To compute the distributions in Eq.~\ref{eq:smalltree},
  we take the tree-based approach described in Fig.~\ref{fig:belieftree}. The
  root of the tree is the belief state $\beta_k(s_1, s_{k}|y^k,\sigma^{k-1})$.
  By iteratively applying Eqs.~\ref{eq:bayesupdate},\ref{eq:transitionupdate}
  for the various choices of $y_{k+1}^{k+g}, \sigma_{k}^{k+g-1}$, we can
  compute the desired distributions in Eq.~\ref{eq:smalltree}. 
  The cost function can be
  written
  \begin{align}
    \min_{\sigma_{k}^{k+g-1}}\,&H\left( S_1 \middle| y^k, \sigma^{k-1}, Y_{k+1}^{k+g}
    \sigma_{k}^{k+g-1} \right)
    =\nonumber\\ &\min_{\sigma_{k}}\mathbb{E}_{Y_{k+1}|\sigma^{k}}\left(
    \cdots\min_{\sigma_{k+g-1}}\mathbb{E}_{Y_{k+g}|\sigma^{k+g-1}}\left(H\left(
    S_1\middle|y^k, \sigma^{k-1}, Y_{k+1}^{k+g} \sigma_{k}^{k+g-1}
        \right)  \right) \cdots\right),\label{eq:iterminentropy}
  \end{align}
so a method similar to that used to solve the Bellman Equation can be applied to obtain the solution to
Eq.~\ref{eq:iterminentropy}.

  After applying $\sigma_{k}$ and observing $y_{k+1}$, we need to update the
  tree of belief states. Instead of recomputing the whole tree, we take
  a dynamic programming approach and leverage the fact that we have already
  computed some of the possible future belief states. 
  Rather than recomputing the new root of the tree, we can simply use the
  already computed $\beta_{k+1}(s_1, s_{k+1}|y^{k+1}, \sigma^{k})$ as the new
  root of the tree. All but the last step of the possible future paths have
  also already been computed, so we can just apply Eqs.~\ref{eq:transitionupdate},\ref{eq:bayesupdate}
  to the distributions $ \Prob\left( s_1, s_{k+g} \middle| y^{k}, \sigma^{k-1},
  y_{k+1}^{k+g}, \sigma_{k}^{k+g-1}\right)$ to
    obtain the new belief states 
    \[\Prob\left( s_1, s_{k+g+1} \middle| y^{k+1},
    \sigma^{k}, y_{k+2}^{k+g+1}, \sigma_{k+1}^{k+g}\right).
  \]
  
  \paragraph*{Reduction to POMDP}
  \label{sec:pomdp}

Although we were able to compute the optimal policy for the examples considered
in this paper, in general it requires a very large computation. For the general
case, it is useful to apply the existing framework of partially
observable Markov decision processes (POMDPs)~\cite{Astrom65}, to leverage
existing approximate algorithms (e.g.~\cite{kurniawati2008sarsop}). 
A POMDP is a generalization of an HMM that includes output-dependent actions,
and includes a reward that is given at each step. The goal of policy planning
in POMDPs is to maximize this reward.

To use the framework of POMDPs, it is necessary to express the reward, in our
case the measurement fidelity, as a sum of rewards at each step. We now sketch
how to adjust the model to accomplish this.
We expand the state space to keep track of the initial state  and the number of steps. The new state space
is  $\cS\times \cL\times \{1,\ldots,n\}$. The transition and output
matrices are redefined accordingly. The action at each step is either a permutation or, for the last step, a decision action that is the estimate of the initial state.
Possible decision actions are  in one-to-one correspondence with $\cL$, so the set
of possible actions is the union of the set of allowed permutations and $\cL$.
The reward at each step is  $0$ if the step number
encoded in the extended state is not $n$ or the  decision action does not correspond to the initial state. If the step number is $n$
and the decision action agrees with the initial state, the reward is $1$.
With this definition, the expectation of the sum of the rewards at each
step is the measurement fidelity.

\section{Implementation and simulation}
\label{sec:impsim}

The code used in our simulations of HMMs with permutations
is available on Github~\cite{github}. The code uses the \textsc{pyro}
package~\cite{bingham2019pyro} and the underlying
PyTorch~\cite{NEURIPS2019_9015} package for computation of relevant probabilities. 

For the examples given in the main text, we computed
 the measurement infidelity numerically without resorting to
  Monte Carlo techniques.
  In particular,  for a given number $n$ of steps, 
  we considered all of the $|\cY|^{n}$ possible output sequences.
  For each such sequence $y^{n}$, we computed the probability that it occurs and
  the  probability of incorrectly identifying the initial state conditional on $y^{n}$.
  The measurement infidelity is obtained by summing the product of these two probabilities
  over $\cY^{n}$.
  The number of possible data sequences is exponential in  $n$,  which limits the number of steps for which it is possible
  to avoid Monte Carlo sampling.   Computation of the probability of an outcome
  sequence and of the posterior initial state distribution used the forward-backward
  algorithm~\cite{Rabiner1989}, which is built into the packages we
  used.

\begin{acknowledgments}
  This work includes contributions of the National Institute of
  Standards and Technology, which are not subject to U.S. copyright.
  The use of trade, product and software names is for informational
  purposes only and does not imply endorsement or recommendation by
  the U.S. government.
  S. Geller acknowledges support from the Professional Research Experience Program (PREP) operated jointly by NIST and the University of Colorado.
  D. C. C. acknowledges support from a National Research Council postdoctoral fellowship.
  E. K. thanks Dietrich Leibfried for introducing them
  in the early days of ion trap quantum computing to the idea
  of adaptively chosen pulses for improving measurement fidelity.  We
  thank Zachary Sunberg for discussions on the POMDP
  formalism. We thank Giorgio Zarantonello for 
    computations involving the transition rates in $^9\text{Be}^+$. We thank
    Ting Rei Tan, Mohammad Alhejji, Alexander Kwiatkowski, Arik Avagyan,
  Akira Kyle, and Stephen Erickson for helpful suggestions and comments.
\end{acknowledgments}

\bibliography{perm_hmm_paper}
\end{document}